\documentclass[fleqn,twoside,twocolumn,nofootinbib,showkeys]{revtex4} 

\usepackage{graphicx}

\begin{document}
\title[Transition to chaos in the kinetic model]
{Transition to Chaos in the Kinetic Model of Cellulose Hydrolysis Under Enzyme Biosynthesis Control}%

\def\bitp{Bogolyubov Institute for Theoretical Physics, Nat. Acad. of Sci. of Ukraine}
\def\bitpaddr{14b, Metrolohichna Str., Kiev 03680, Ukraine}

\author{A.S. Zhokhin}
\affiliation{\bitp}
\address{\bitpaddr}
\email{aszhokhin@bitp.kiev.ua}

\author{V.P. Gachok}
\affiliation{\bitp}
\address{\bitpaddr}
\email{vgachok@bitp.kiev.ua}

\begin{abstract}
In the paper the kinetic model of the biochemical process of cellulose
hydrolysis with cell application is presented. The model includes enzyme
biosynthesis control and is open conditions it represents the dynamical system
in the preturbulent regime. The limit cycle and its five consequence 
bifurcations of the doubling-period type are found. Also the limit regime of the
system - the strange attractor - is presented.
\end{abstract}

\maketitle

\section{Introduction}

In the paper we present the dynamical system which has a direct interpretation in
the biochemical process of the cellulose hydrolysis [1-3]. The factors present
in the system such as the biosynthesis of cellobiase, the repression of
biosynthesis by glucose, the cell lysis and the conditions of the input of 
initial substrate and the output of cells and connected to it the inactivation
of cellobiase create the living conditions with complicated regimes in the kinetics 
of the system. The constructed biosystem is found to develop periodic
oscillatory regimes including the chaotic regime, the strange attractor.
In chemical kinetics the dynamical system with a strange attractor
was studies by R\"osler \cite{4}.

\section{The Description of the Model}

\begin{figure}%
\vskip1mm
\includegraphics[width=5cm]{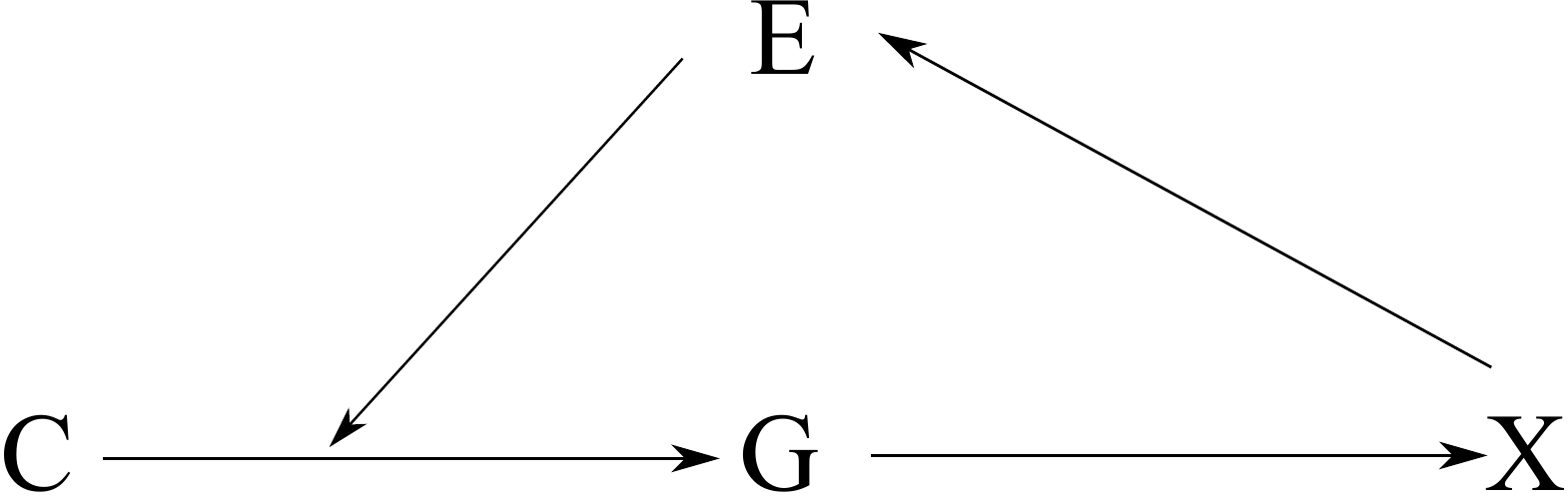}
\vskip-3mm\caption{General reaction scheme of the processes}\label{fig:1}
\end{figure}

We consider the kinetics of hydrolysis of cellulose $C$ under action of the
enzyme $E$ whose activity is controlled by the biosynthesis in the 
microorganisms $X$. The end product is glucose $G$.
This kinetics is described by the following 4-dimensional system of differential
equations being the simplification of the cellulose hydrolysis model proposed in \cite{1,2,3}:

\vspace*{20mm}

\begin{equation}\label{eq1}
 \frac{dC}{dt}=a - l \frac{E}{(1+E)}\frac{C}{(1+C+G)},
\end{equation}\vspace*{-1mm}
\begin{equation}\label{eq2}
 \frac{dG}{dt}=2l \frac{E}{(1+E)}\frac{C}{(1+C+G)}-m \frac{GX}{(1+X+G)},
\end{equation}\vspace*{-1mm}
\begin{equation}\label{eq3}
 \frac{dX}{dt}=m_1 \frac{GX}{(1+X+G)}-m_0X,
\end{equation}\vspace*{-1mm}
\begin{equation}\label{eq4}
 \frac{dE}{dt}=E_{0}\frac{C}{(1+C)}\frac{N}{(N+G)}-eE,
\end{equation}\vspace*{7mm}

\begin{figure}%
\vskip1mm
\includegraphics[width=7.5cm]{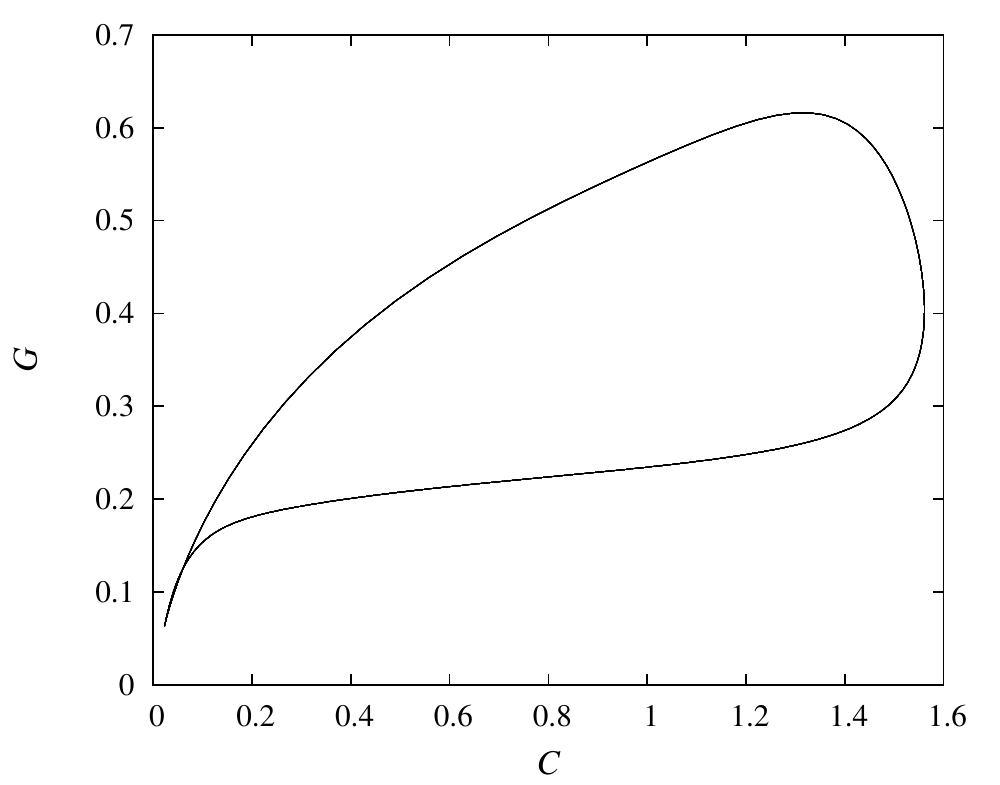}
\vskip-3mm\caption{$f=43$, Limit cycle. The period is $T$=2.03}

\label{fig:2}\vspace*{-2mm}
\end{figure}

\begin{figure}%
\vskip1mm
\includegraphics[width=7.5cm]{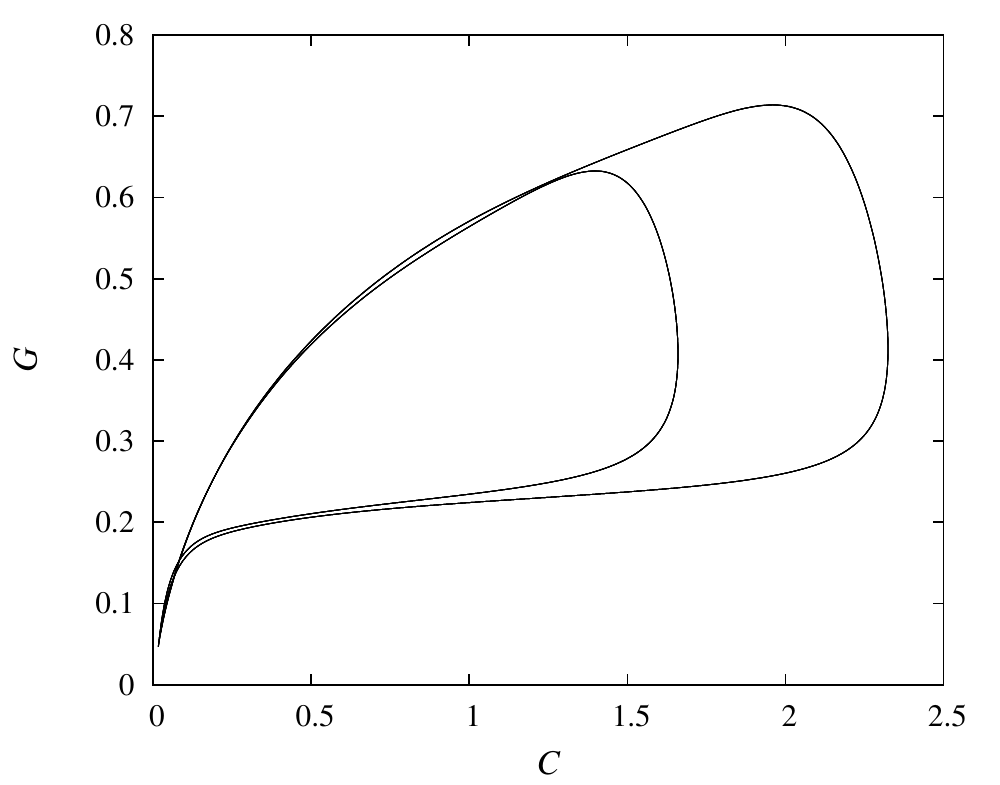}
\vskip-3mm\caption{$f=43.5$, Limit cycle. The period is $T$=4.94}

\label{fig:3}\vspace*{-2mm}
\end{figure}

\begin{figure}%
\vskip1mm
\includegraphics[width=7.5cm]{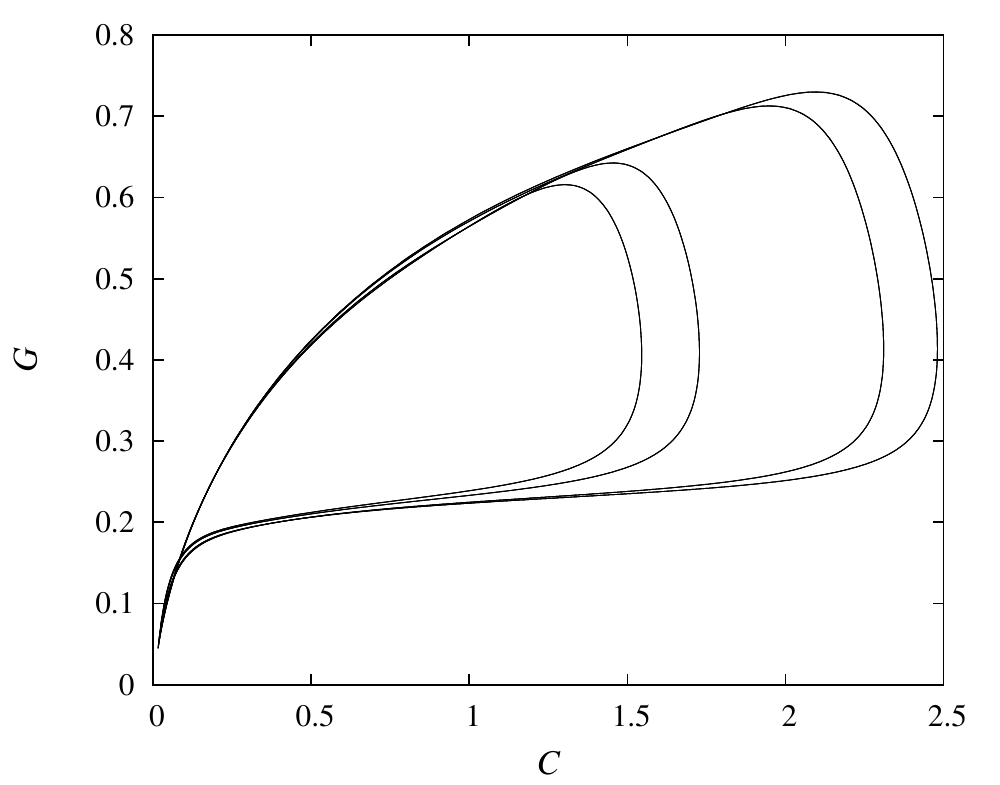}
\vskip-3mm\caption{$f=43.55$, Limit cycle. The period is $T$=10.3}

\label{fig:4}\vspace*{-2mm}
\end{figure}

\begin{figure}%
\vskip1mm
\includegraphics[width=7.5cm]{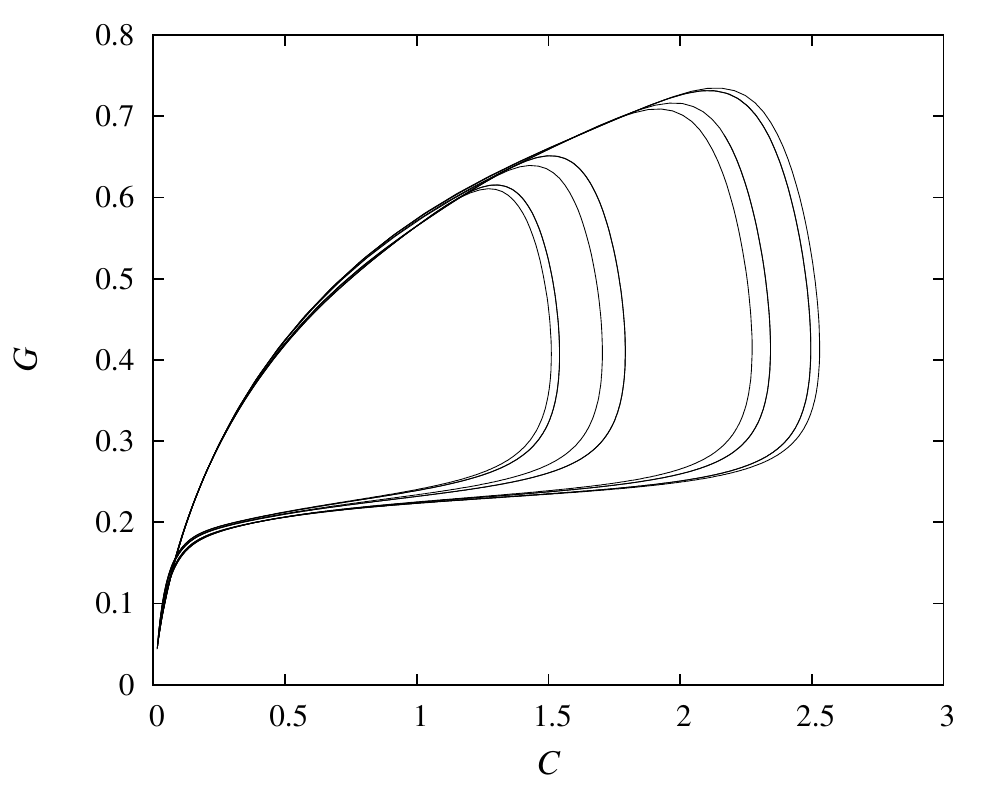}
\vskip-3mm\caption{$f=43.565$, Limit cycle. The period is $T$=20.8}

\label{fig:5}\vspace*{-2mm}
\end{figure}

\begin{figure}%
\vskip1mm
\includegraphics[width=7.5cm]{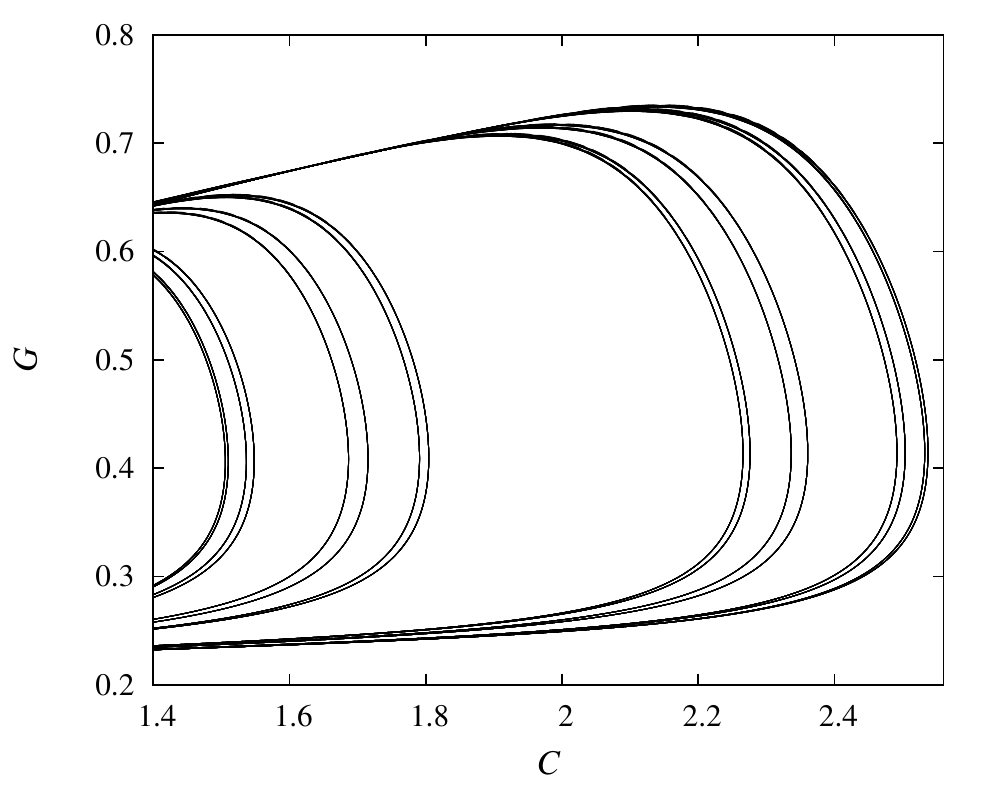}
\vskip-3mm\caption{cycle. The period is $T$=41.1}

\label{fig:6}\vspace*{-2mm}
\end{figure}

\begin{figure}%
\vskip1mm
\includegraphics[width=7.5cm]{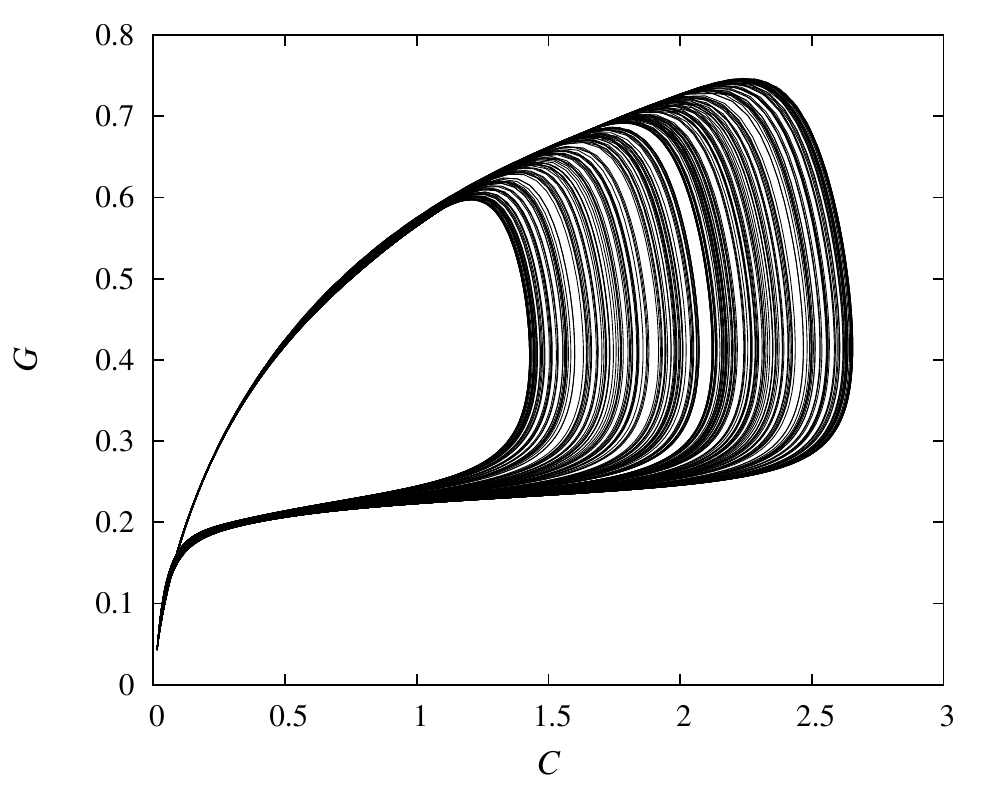}
\vskip-3mm\caption{$f=43.6$, Strange attractor}

\label{fig:7}\vspace*{-2mm}
\end{figure}

where $C,G,X,E$ are the dimensionless concentrations of the corresponding reagents.
The parameter $a$, according to the first equation, represents the input of the
cellulose $G$ in the system and terms $fX$ and $eE$, according to the third and the
fourth equations, characterize the output of the microorganisms $X$ and the
enzyme $E$ (the dissipation). The system shows that the growth of substrate $C$ 
includes the enzyme biosynthesis $E$ which, consequently, leads to the production 
of $G$ and the growth of $X$. In turn the growth of $X$ increases the biosynthesis of $E$.
At large $E$ and $X$ the level of $C$ and $G$ decreases. Further the output leads to
small concentrations of $E$ and $X$ and again due to the input $C$ they start to
increase and the process repeats. The existence of these feedback controls
leads to the complicated oscillations.

\section{Computer Results}

We fix the parameters, except $f$, as follows:
$a=1,l=200,m=400,m_1=230,$
$E_0=1100, N=0.05, e=20.$
We investigate the phase portrait of the system changing the parameter $f$ in the
bounds 43 $\le f \le 43.6.$
In this case the system has the stationary  focus in the phase space.
However, the stable focus is not a global attractor. The system with the above
$f$ has one more attractor. It was studied by computer simulations.
The Cauchy data  were: $G_0=0.9, G_0=0.24,X_0=0.64,E_0=0.8.$
The period was found with the application of the Poincar\'e map.
This initial state of the system develops to the limit cycle.
In growth of $f$ we found that at the points
$f_1=43.39, f_2=43.53, f_3=43.562, f_4=43.5667, f_5=43.5677$,
the double period cycle occurs. The following increase of $f$ leads to the
chaotic regime, the strange attractor. Thus we have constructed the scenario of 
the following bifurcations and its limit, the strange attractor.
The results are represented in the phase plane $(C,G)$ in the Fig.~\ref{fig:2}~\
- ~\ref{fig:7}.


\vspace*{2mm}
\begin{thebibliography}{99}

\bibitem{1} V.P. Gachok, ITP-81-102R, Kiev (1981)\vspace*{1mm}
\bibitem{2} V.P. Gachok, ITP-83-33E, Kiev (1983)\vspace*{1mm}
\bibitem{3} V.P. Gachok, A.S. Zhokhin, ITP-83-54R, Kiev (1983)\vspace*{1mm}
\bibitem{4} O. R\"ossler, Bull.Math.Biol.,v39 (1977) 275\vspace*{1mm}
\bibitem{5} Th. Schulmeister, studia biophysica, v 105 (1985) 5\vspace*{1mm}


\begin{flushright}
\end{flushright}
\end{thebibliography}
\end{document}